\documentclass[aps,prb,twocolumn,superscriptaddress,showpacs]{revtex4-1}
\usepackage[dvips]{graphicx}
\begin{document}

\title{Hysteretic phenomena in a 2DEG in quantum Hall effect regime studied in a transport experiment}

\author{M.~V.~Budantsev}
\email[]{budants@isp.nsc.ru}
\affiliation{A. V. Rzhanov Institute of Semiconductor Physics of SB RAS, Novosibirsk, Russia}
\affiliation{Novosibirsk State University, Russia}

\author{D.~A.~Pokhabov}
\author{A.~G.~Pogosov}
\author{E.~Yu.~Zhdanov}
\affiliation{A. V. Rzhanov Institute of Semiconductor Physics of SB RAS, Novosibirsk, Russia}
\affiliation{Novosibirsk State University, Russia}

\author{A.~K.~Bakarov}
\author{A.~I.~Toropov}
\affiliation{A. V. Rzhanov Institute of Semiconductor Physics of SB RAS, Novosibirsk, Russia}

\date{\today}

\begin{abstract}
We investigated experimentally non-equilibrium state of a two-dimensional electron gas (2DEG) in the quantum Hall effect (QHE) regime, studying the hysteresis of magnetoresistance of a 2DEG with a constriction. The large amplitude of the hysteresis enabled us to make the consistent phenomenological description of the hysteresis. We studied the dependence on the magnetic field sweep prehistory (minor loop measurements), recovered the anhysteretic curve, and studied the time dependence of the magnetoresistance. We showed that the hysteresis of magnetoresistance of a 2DEG in the QHE regime has significant phenomenological similarities with the hysteresis of magnetization of ferromagnetic materials, showing multistability, jumps of relaxation, and having the anhysteretic curve. Nevertheless, we revealed the crucial difference, manifested itself in an unusual inverted (anti-coercive) behavior of the magnetoresistance hysteresis. The time relaxation of the hysteresis has fast and slow regimes, similar to that of non-equilibrium magnetization of a 2DEG in QHE regime pointing to their common origin. We studied the dependence of the hysteresis loop area on the lithographic width of the constriction and found the threshold value of width $\sim$1.35~$\mu$m beyond which the hysteresis is not observed. This points to the edge nature of the non-equilibrium currents (NECs) and allows us to determine the width of the NECs area ($\sim$0.5~$\mu$m). We suggest the qualitative picture of the observed hysteresis, based on non-equilibrium redistribution of the electrons among the Landau level states and assuming huge imbalance between the population of bulk and edge electronic states.
\end{abstract}

\maketitle

\section{Introduction\label{Introduction}}
A two-dimensional electron gas (2DEG) in the quantum Hall effect (QHE) regime demonstrates a number of intriguing phenomena that are not fully explained up to date. Among them are non-equilibrium phenomena manifested in hysteresis of sheet electron density, \cite{Pudalov1984, Dolgopolov1992, Dolgopolov1993} magnetization (see for review Ref.[\onlinecite{Usher2009}]) and electrochemical potential \cite{Huels2004, Klaffs2004} as a function of magnetic field and gate voltage (for gated structures). The hysteresis was observed in an ordinary single-layer 2DEG in nonmagnetic materials such as Si MOSFETs and AlGaAs/GaAs heterostructures at integer and fractional filling factors corresponding to a longitudinal resistance ($R_\mathrm{L}$) vanishing.

It was found that the hysteretic variations of magnetization \cite{Usher2009} and electrochemical potential \cite{Huels2004, Klaffs2004} exceed equilibrium values by a factor of 10-60. In contrast the conventional electron transport measurements in QHE regime with vanishing $R_\mathrm{L}$ are not sensitive to the hysteresis, though, the question of a probable indirect influence of the non-equilibrium on the transport properties of 2DEG still remains open.

Recently in a number of papers it was demonstrated that creation of a constriction in a macroscopic bath of 2DEG allows to reveal the hysteretic behavior in magnetoresistance \cite{Budantsev2007, Budantsev2009, Budantsev22009, Budantsev2011, Pioro-Ladriere2006, Smith2011}.

It was shown that at low temperatures (lower than 100~mK) the hysteresis relaxes over a day \cite{Pioro-Ladriere2006} which is far beyond the time usually spent on a conventional measurement (a magnetic field sweep) in the QHE regime. In terms of usual magnetic hysteresis description \cite{Bertotti1998} this type of hysteresis is considered as time-independent.

The most common explanation of the hysteretic phenomenon is based on non-equilibrium currents (NECs) slowly relaxing providing $R_\mathrm{L}$ vanishing. But there are no unambiguous models in the literature, adequately describing this phenomenon. The detailed consideration of NECs induction is proposed in the Ref.~ [\onlinecite{Shikin2002}], but it concerns only the case of weak non-equilibrium. Ruhe et al. \cite{Ruhe2009} studied the case of time-dependent hysteresis. The most developed model up to now \cite{Matthews2004} is based on the assumption of redistribution of ``frozen'' charges over the whole bulk area of the 2DEG \cite{Dyakonov1991}. These charges give rise to the radial (for the disk geometry) electric field that in turn gives rise to azimuthal drift magnetization current. The magnetization maximum is determined by the QHE breakdown current, calculated in the frames of quasi-elastic inter-Landau level scattering (QUILLS) approach.\cite{Eaves1986} The mentioned model correctly describes experimentally observed linear temperature suppression of the hysteresis amplitude observed both in magnetization \cite{Matthews2004} and constriction conductance.\cite{Budantsev2007, Budantsev2009} But the magnetization maximum calculated in the model is less than experimentally obtained value by a factor of 4. Moreover, the model predicts bulk NECs distribution, which, to our opinion, contradicts with experimental data of Klaffs et. al. \cite{Klaffs2004}

In the present work we study the hysteresis of the magnetoresistance of the conventional 2DEG with constriction, placed in it. It was found that at certain critical width of the constriction the hysteresis of the magnetoresistance vanishes. This observation strongly suggests that NECs are localized near the edge of the sample.

A phenomenological comparison of the 2DEG magnetoresistance hysteresis in the QHE regime with the hysteresis of magnetization of ordinary ferromagnetic materials is performed. Particularly, we discuss the dependence on the magnetic field sweep prehistory for both cases, studying minor loops, anhysteretic curves and time relaxations. Mostly we observe a similar behavior. However significant difference has been found manifested itself in unusual advancing (anti-coercive) response (2DEG magnetoresistance) to external parameter (magnetic field) change, which is just opposite to the behavior of ferromagnetic materials.

We compare our results with the hysteresis of magnetoresistance in systems with pseudo-spin degree of freedom --- so-called ``quantum Hall ferromagnets'' \cite{Piazza1999, DePoortere2000, Jungwirth2001}. In some cases (see e.g. Ref.~[\onlinecite{Piazza1999}]) analogical anti-coercive behavior is found. This phenomenological similarity suggests that there are similar mechanisms underlying the origin of the hysteresis both in ordinary 2DEG and in QH ferromagnets.

In addition to the phenomenological description, we provide a qualitative picture of a spatial distribution of NECs and electrochemical potential in the sample.

The paper is organized in the following way. In Sec.~\ref{Experimental details} we describe the experimental detail. In Sec.~\ref{Experimental results} we present the experimental data of following measurements: the magnetoresistance of the 2DEG with the constriction in the QHE regime, minor loop, anhysteretic curve, time relaxation dependence of the magnetoresistance, and the dependence of the magnetoresistance hysteresis loop area on the lithographic width of the constriction and discuss them. In Sec.~\ref{Discussion} we develop a detailed qualitative picture of spatial distribution of NECs in a sample, explaining the hysteresis of the magnetoresistance of the constriction as well as the non-equilibrium magnetization of a 2DEG in the QHE regime. We discuss inverted (anti-coercive) behavior of the hysteresis and compare the hysteresis of the magnetoresistance with the hysteresis of the ferromagnetic materials magnetization. We conclude in Sec.~\ref{Conclusions}.

\section{Experimental details\label{Experimental details}}
\begin{figure}
\includegraphics[width=.49\textwidth]{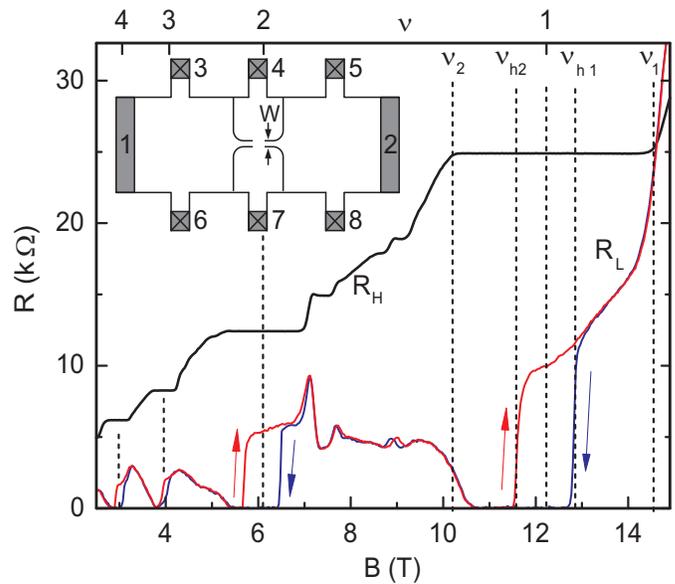}
\caption{The magnetoresistance of the samples of the type I. At the centers of the plateaux of macroscopic 2DEG Hall resistance $R_\mathrm{H}~=~U_{36}/I_{12}$ at odd and even integer filling factors a
giant hysteresis of longitudinal magnetoresistance $R_\mathrm{L}~=~U_{35}/I_{12}$ is observed. The geometry of the sample is shown in the inset.\label{fig1}}
\end{figure}

Experimental samples were fabricated from GaAs/AlGaAs heterojunctions with 2DEG of two types (type I and type II) grown in different cycles of molecular beam epitaxy. The electron mobility $\mu$ and spacer layer thickness $d$ are $\mu$=0.6$\div$0.8$\times$10$^{6}$~cm$^{2}$/V$\cdot$s and $d=400$~{\AA} in the type I sample and  $\mu$=0.8$\div$1.0$\times$10$^{6}$~cm$^{2}$/V$\cdot$s and $d=300$~{\AA} in the type II sample. The electron densities of the macroscopic 2DEGs of both types are n$_\mathrm{s}$=3$\times$10$^{11}$~cm$^{-2}$ at the temperature of 4.2~K. Hall bars of the dimensions $W$$\times$$L$=50$\times$100~$\mu$m$^{2}$ have been created on the surface of the heterostructures by means of photolithography. In the middle part of the Hall bars constrictions of the effective width 0.6$\div$0.8\,$\mu$m and the length 3~$\mu$m have been created by means of electron beam lithography (see the inset of Fig.~\ref{fig1}). The samples of the both type allow to measure longitudinal magnetoresistance $R_\mathrm{L}=U_{35}/I_{12}$ and Hall magnetoresistance of macroscopic 2DEG $R_\mathrm{H}=U_{36}/I_{12}$. The measurements are carried out in the linear response regime on the alternating current of the magnitude 1$\div$10~nA and the frequency 7~Hz at the temperature of 60~mK. The magnetic field covering a range of 0$\div$15\,T was oriented perpendicular to the 2DEG.
\section{Experimental results\label{Experimental results}}
First, we have studied the type I samples. A giant hysteresis of longitudinal magnetoresistance $R_\mathrm{L}$ of the constriction with respect to magnetic field sweep direction is observed in magnetic fields corresponding to centers of the plateau of macroscopic 2DEG Hall resistance $R_\mathrm{H}$ at both odd and even integer filling factors (Fig.~\ref{fig1}). The effect is independent on a direction of magnetic field.
The most pronounced hysteresis loop is observed near the filling factor $\nu=1$ in the range of filling factors between $\nu_\mathrm{h1}=0.95$ and $\nu_\mathrm{h2}=1.07$ on the background of macroscopic 2DEG Hall plateau observed in the range of filling factors between $\nu_1=0.85$ and $\nu_2=1.22$ (Fig.~\ref{fig1}). The hysteresis has the amplitude 10\,k$\Omega$ amounting to 100$\%$ of measuring signal and the width width of the hysteresis is 1.2~T at filling factor $\nu=1$. Drastic step-like rise of the magnetoresistance from zero up to 10~$k\Omega$ at up-sweep of the magnetic field and sharp drop down to zero at down-sweep of the magnetic field allows us to consider this behavior as the magneto-induced breakdown of QHE.

 The large amplitude of the hysteresis made it possible to carry out a detailed study. It has been found that the response of the system (magnetoresistance) outpaces the external excitation (magnetic field change) rather than retards from it, unlike the case of ferromagnetic magnetization. In mathematical terms the hysteresis loop is negatively oriented curve, while in ferromagnetic materials the hysteresis loop is positively oriented. One can see that the magnetoresistance dramatically changes at the entrance to the area of hysteresis, that is $|dR/dB|$ tends to a huge value, and it remains almost unchanged when leaving the area. In the case of retarded behavior the magnetoresistance would practically not be changed at the entrance to the area of hysteresis, that is $|dR/dB|$ would be close to zero. In this sense the observed hysteresis is abnormally inverted having a negative coercivity in contrast to the ferromagnetic magnetization.

\begin{figure}
\includegraphics[width=.49\textwidth]{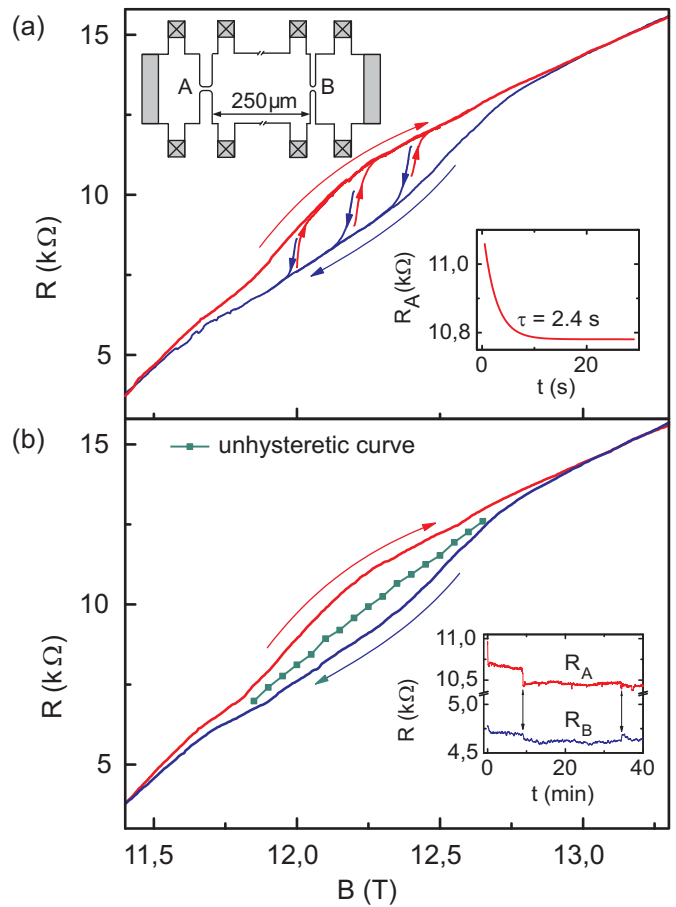}
\caption{
The hysteresis of the magnetoresistance of the II type sample at the filling factor $\nu=1$. (a) Major and minor loops. (b) Anhysteretic curve.  Arrows indicate the magnetic field sweep direction. The sweep-rate is 0.05~T/min. The geometry of the sample and relaxation are shown in the insets. One can see the fast relaxation regime with $\tau$=2.4~s. The relaxation is occurred by jumps simultaneously in both structures.\label{fig2}}
\end{figure}

The magnetoresistance of the type II sample also demonstrates the hysteresis of magnetoresistance (Fig.~\ref{fig2}) in the range of filling factors between $\nu_\mathrm{h1}=0.96$ and $\nu_\mathrm{h2}=1.06$ on the background of macroscopic 2DEG Hall plateau observed in the range of filling factors between $\nu_1=0.88$ and $\nu_2=1.16$. One can see that the hysteresis of magnetoresistance of the type II samples is observed at the same range of filling factors as in the type I samples but it does not demonstrate a giant amplitude and drastic jumps. This distinction is probably resulted from the differences in heterostructures parameters and needs to be discussed separately. Nevertheless, we succeeded to reveal inverted (anti-coercive) behavior of the magnetoresistance hysteresis of the type II sample studying the form of minor hysteretic loops. One can see that a change of the magnetic field sweep direction inside the major hysteretic loop leads to dramatic change of the magnetoresistance, i.e. the derivative $|dR/dB|$ tends to a huge value (Fig.~\ref{fig2}(a)), while the retarding behavior would correspond to zero $|dR/dB|$ (as in the case of minor loops of ferromagnetic materials, where $|dM/dH|=0$). Moreover the analysis of the minor loops allows us to conclude that there is a large number of sample states inside the major hysteresis loop, and therefore the sample state is multistable.

It is generally accepted that long-lived NECs are induced on the background of Hall plateaus where $\sigma_{xx}$ is vanishing. Our measurements shows that the range of existence of the hysteresis $\Delta\nu_\mathrm{h}$ is approximately 3 times less than the Hall plateau width $\Delta\nu$: $\Delta\nu_\mathrm{h}$/$\Delta\nu$$\approx$1/3. That is the condition $\sigma_{xx}\approx0$ is necessary but not sufficient to observe the hysteresis.

To perform consistent phenomenological comparison of the 2DEG magnetoresistance hysteresis in QHE regime with the hysteresis of magnetization of ferromagnetic materials we have measured the anhysteretic curves (Fig.~\ref{fig2}(b)). Each experimental point $R_i(B_i)$  has been obtained by cycling the magnetic field around $B_i$ with a decreasing amplitude. The form of the anhysteretic curve and its position inside the loop are phenomenologically similar to that of ferromagnetic materials.

We have studied the time relaxation of the hysteresis, measuring the time dependence of the magnetoresistances of two conducting channels situated in the same Hall bar and separated by macroscopic 2DEG reservoir of the length of 250~$\mu$m and the width of 50~$\mu$m (see the top inset of Fig.~\ref{fig2}(a)). The dependencies are shown in the insets of Fig.~\ref{fig2}(a) and (b). The curves have been obtained after stopping the up-sweep of the magnetic field at the value 12.25~T, corresponding to the filling factor $\nu=1$ where the hysteresis has the maximum amplitude.

It has been found that time relaxation has two phases: the fast initial one followed by slow phase. After stopping the magnetic field sweep we have observed the fast exponential relaxation by relatively small value (less than 25$\%$) with relaxation time 2.4~s (see the bottom inset of Fig.~\ref{fig2}(a)). Such relaxation time is too big to be explained by spin-orbital coupling mechanism of relaxation \cite{Muller1992, Khaetskii1992} and resembles the nuclear relaxation time in GaAs which is about 30~s.\cite{Dixon1997, Devyatov2004} It should be noted that the non-equilibrium magnetization relaxes in the similar way.\cite{Kershaw2007} It also has regime of initial exponential decay followed by a much slower power-law decay, however the relaxation time in the first regime is longer ($\sim$20~s) than obtained in our work. In Ref.~[\onlinecite{Matthews2004}] it has been suggested that magnetic field change results in QHE breakdown accompanied by electron transitions between adjacent Landau levels with opposite spin. The observed fast relaxation can be explained by such inter-Landau levels scattering and brings the system to a local (not global) minimum of energy after stopping the magnetic field sweep. According to Ref.~[\onlinecite{Matthews2004}] such scattering results from QUILLS process \cite{Eaves1986} and takes place in a part of the sample where the maximum electric field is reached. In Refs.~[\onlinecite{Matthews2004},~\onlinecite{Dyakonov1991}] it is suggested that maximum electric field is reached in the bulk of a sample. However our results are sensitive to the sample edge (see Section~\ref{Discussion}) since the magnetoresistance is defined by the edge states filling. This allows us to conclude that QHE breakdown takes place at the sample edge.

Further relaxation is much slower. During the next 40~min the resistances of both structures relax by a small value ($\sim$20$\%$). Moreover the most notable changes are occurred by sudden jumps simultaneously in both remote structures (see the inset of Fig.~\ref{fig2}(b)). Taking into account that the structures are separated by macroscopic 2DEG reservoir of the length of 250~$\mu$m and the width of 50~$\mu$m it can be concluded that these jumps are caused by relaxation process in the macroscopic 2DEG reservoir. This observation once again confirms the conclusion that the narrow conductive channel is a tool for the study of the non-equilibrium phenomena taking place in a macroscopic 2DEG.\cite{Budantsev2007} The observed stepwise relaxation is phenologically similar to Barkhausen jumps observed in the ferromagnetic materials and originating from spin domain structure transformations.\cite{Bertotti1998}

A formation of a spatial spin polarization in a single quantum wire has been considered in Ref.~[\onlinecite{Ihnatsenka2007}] and it has been shown that this would result in hysteresis in electron transport measurements. However, all the calculated characteristics demonstrate the conventional coercive behavior. Consequently, the spin mechanism is not relevant to our results. Moreover, earlier it have been shown that the hysteresis of magnetoresistance is independent on the in-plain component of magnetic field.\cite{Budantsev2009} It allowed us to conclude that the hysteretic effect is not related to the electron spin.

The measurements of non-equilibrium magnetization of a 2DEG also show stepwise relaxation.\cite{Smith2011} It should be noted that transport measurements give a valuable complementary information about the non-equilibrium state of a 2DEG. Fact is that the magnetization is an integral characteristic determined by the sum of all the magnetic moments, which are caused by all magnetization currents in a sample, while the magnetoresistance is sensitive to a local electrochemical potential of edge states. Particularly in this paper we succeed to observe the correlation between the relaxation of magnetoresistance of two remote constrictions and have established that NECs are localized near the edge.
\begin{figure}
\includegraphics[width=.49\textwidth]{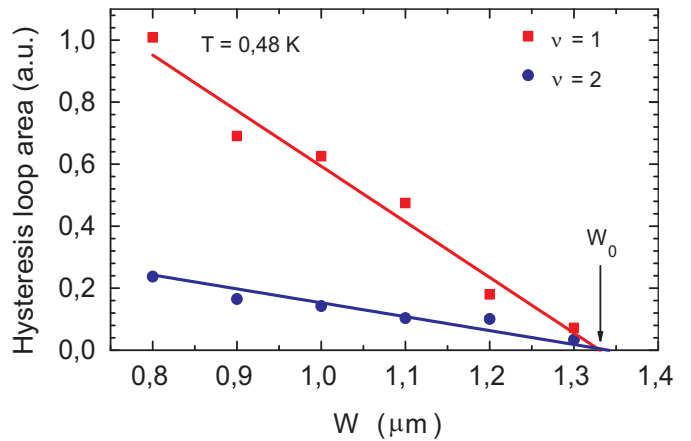}
\caption{The dependence of the hysteresis loop area on the constriction width at the filling factors $\nu$=1 and 2 at temperature 0.48~K.\label{fig3}}
\end{figure}

To determine the width of spatial localization of NECs the dependence of magnetoresistance hysteresis loop area on the lithographic width of the constriction has been experimentally studied. For this purpose we used the series of samples fabricated from the same heterostructure. The samples were the Hall bars with constrictions of different widths. The constrictions of widths from 0.8~$\mu$m to 1.3~$\mu$m have been fabricated by means of electron beam lithography in the central part of Hall bars. All the constrictions have low resistance in zero magnetic field (0.2$\div$0.8~k$\Omega$) and demonstrate QHE in high magnetic fields. At the temperature of 0.48~K the hysteresis of magnetoresistance have been observed at the filling factors $\nu$=1, 2 and 4 for all the samples. Fig.~\ref{fig3} shows the hysteresis loop area as a function of the constriction width at the filling factors $\nu$=1 and 2. Experimental points are well fitted by linear functions. One can see that the hysteresis loop area decreases with the constriction width and vanishes at certain critical width $W_{0}$. Moreover these critical widths coincide for both filling factors and are about $W_{0}\approx$1.35$\mu$m. Assuming that the critical lithographic width comprises of two counter-propagating NECs of the width $W_\mathrm{NEC}$ and two depletion regions of the width $W_\mathrm{depl}\approx$0.2$\mu$m: $W_{0}=2W_\mathrm{NEC}+2W_\mathrm{depl}$, we obtain $W_\mathrm{NEC}\approx$0.5$\mu$m is comparable with the depletion width. The edge character of NECs has been also reported in the Ref.~[\onlinecite{Klaffs2004}] devoted to the study of spatial distribution of the non-equilibrium electrostatic potential on a 2DEG surface, where it has been shown that the most significant changes of potential take place close to the edge.

The observed hysteresis of magnetoresistance exhibits the following properties: (i) it has two phases of  relaxation, fast and slow; (ii) the slow relaxation is occurred by jumps resembling the Barkhausen jumps in ferromagnetics; (iii) it demonstrates inverted anti-coercive behavior; (iv) it shows multistability and has an anhysteretic curve; (v) it can be observed in sufficiently narrow channel thus pointing out its edge character.

\section{Discussion\label{Discussion}}
The existing physical models are able to interpret only separate experiments. Unfortunately there is no any model which would not contradict all the set of experimental data on the non-equilibrium state in a 2DEG in the QHE regime. Analyzing the experimental data obtained in our and other studies\cite{Budantsev2007, Klaffs2004, Matthews2004, Usher2009, Ruhe2009} we suggest such a qualitative picture. We propose the picture of spatial distribution of NECs in a 2DEG and discuss its relation to the experimentally observed non-equilibrium magnetization of the 2DEG. For simplicity, we consider a 2DEG at the filling factor near $\nu=1$, when electrons form the incompressible liquid on the first Landau level and inter Landau level scattering is assumed to be suppressed.

\subsection{Magnetoresistance of constriction\label{Magnetoresistance of constriction}}
The obtained experimental data on the magnetoresistance can be interpreted in the frame of the following physical picture.

\begin{figure}[!]
\includegraphics[width=.49\textwidth]{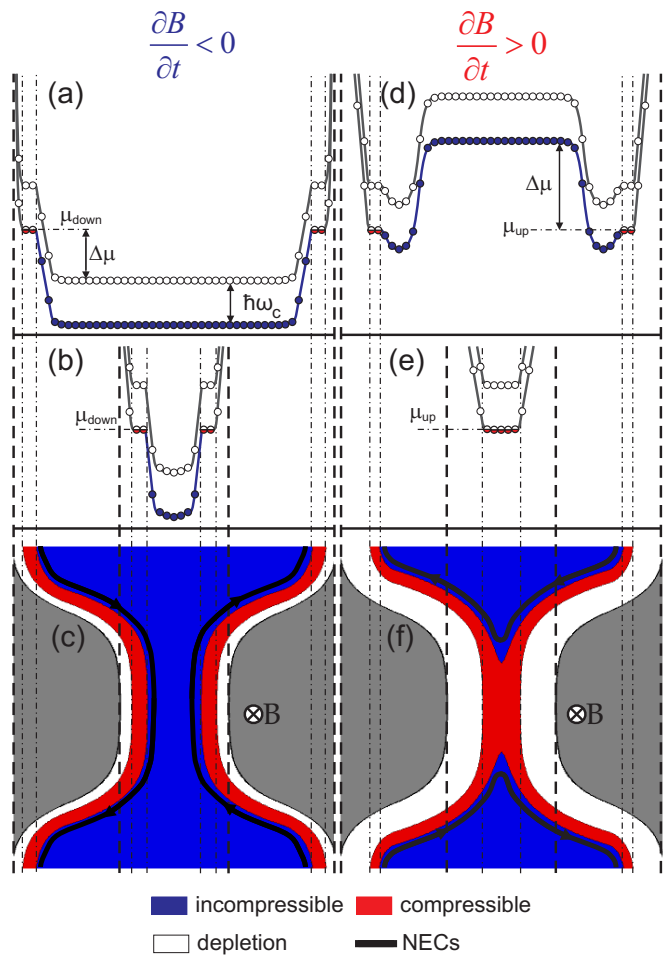}
\caption{Non-equilibrium electron redistribution in the macroscopic 2DEG (a), (d) and in the constriction (b), (e) and top view of the 2DEG with narrowing (c), (f) at up- and down-sweep of the magnetic field correspondingly at the filling factor $\nu$=1. Bulk states in the constriction are occupied (c) or not occupied (f) with incompressible liquid in different magnetic field sweep direction. Arrows designate nonequilibrium currents.\label{fig4}}
\end{figure}

Down-sweep of the magnetic field gives rise to the vortex electric field that induces outflow of the electrons from the bulk to the edge.\cite{Dolgopolov1992, Laughlin1981} In this case the area of the incompressible liquid increases and the edge channels are shifted closer to the lithographic borders of the sample. It means that the counter-propagating edge currents in the constriction become more distant from each other (Fig.~\ref{fig4}(a-c)) than in the case of equilibrium state and the backscattering is suppressed. In the type II samples the magnetoresistance is indeed lower than its equilibrium value on the anhysteretic curve. In the type I samples at filling factor $\nu=1$ the decrease of magnetic field leads to complete suppression of the backscattering in the constriction, manifested in the vanishing of the longitudinal magnetoresistance and the establishment of the QHE regime.

Up-sweep of the magnetic field leads to vortex electric field, which in turn induces outflow of the electrons from the edge to the bulk. In this case the area of the incompressible liquid decreases and the edge channels are shifted away from the lithographic border of the sample. The counter-propagating edge currents in the constriction come closer together (Fig.~\ref{fig4}(d-f)) than in the case of equilibrium state and the backscattering increases. This results in magnetoresistance rise as we have observed in both types of the samples. The magnetoresistance of the type II samples is indeed higher than its equilibrium value on the anhysteretic curve. In the type I samples at the filling factor $\nu=1$ the increase of magnetic field switches the constriction from the QHE regime to the resistive state. In other words, the magnetic field sweep leads to magnetic field induced QHE breakdown.

\subsection{Non-equilibrium magnetization\label{Non-equilibrium magnetization}}
Let us consider how the proposed picture corresponds to the non-equilibrium magnetization measurements. The spatial distribution of the NECs have been discussed in Refs.~[\onlinecite{Usher2009}, \onlinecite{Ruhe2009}, \onlinecite{Matthews2004}]. However, the details of the mentioned models are not fully consistent with the experimental data obtained up to date. In particular, the model proposed in Refs.~[\onlinecite{Usher2009}, \onlinecite{Matthews2004}] assumes that the non-equilibrium charge is redistributed over the whole plane of a 2DEG and consequently the electric field is formed in the bulk. This contradicts experimental data obtained in the present study and discussed in Ref.~[\onlinecite{Klaffs2004}] pointing to the fact that the non-equilibrium charge redistribution takes place only near the edge. In Ref.~[\onlinecite{Ruhe2009}] the spatial distribution of the magnetization currents at the 2DEG edge is qualitatively described (Fig.8 in [\onlinecite{Ruhe2009}]), but the provided picture corresponds rather to equilibrium state because the common Fermi level for the bulk and the edge is introduced, up to which all the electronic states are occupied. Besides in this case the observed magnetization should not exceed the amplitude of the equilibrium dHvA oscillations while the experimental non-equilibrium magnetization is about 20$\div$60 times larger.

\begin{figure}
\includegraphics[width=.49\textwidth]{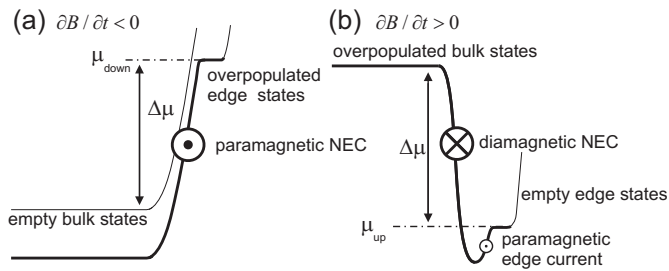}
\caption{The structure of the non-equilibrium edge currents at down- (a) and up-sweep (b) of the magnetic field.\label{fig5}}
\end{figure}

The decrease of magnetic field at the filling factor $\nu=1$  shifts the electrons from the bulk to the edge within the first Landau level. Due to the fact that the area occupied by the bulk states is much larger than that occupied by the edge states one can neglect the changes of the electrostatic potential in the bulk and consider that it remains flat. In this case the most significant changes are occurred near the edge, where a dipolar strip decreasing the electrostatic potential in the bulk and increasing it near the edge is formed. The electrons on the first Landau level populate the edge states up to the level of non-equilibrium electrochemical potential $\mu_{down}$ exceeding the potential of the empty bulk states by the value $\Delta\mu$ (Fig.~\ref{fig5}(a)). In the crossed electric and magnetic fields the electrons drift along the edge forming the magnetization current approximately equal to $e^2/h\cdot\Delta\mu$. This current corresponds to the experimentally observed non-equilibrium paramagnetic magnetization.\cite{Matthews2004, Ruhe2009}

The increase of magnetic field at the filling factor $\nu=1$ leads to more complicated distribution of the magnetization currents. The outflow of the electrons from the edge to the bulk leads to the formation of the dipolar strip near the edge that rises the electrostatic potential in the bulk and forms the potential well near the edge followed by the depletion region. It means that the electrostatic potential is non-monotonic and has a minimum near the edge, caused by interplay of two oppositely directed electric fields: the field of overpopulated bulk and the depletion field (Fig.~\ref{fig5}(b)). On either side of the potential minimum the incompressible electron liquid forms counter-propagating drifting currents  --- giant diamagnetic current flowing closer to the bulk and low paramagnetic current flowing in close proximity to the edge. The presence of the paramagnetic current is confirmed by the experimental fact that the sign of the Hall resistance $R_H$ does not depend on the magnetic field sweep direction in a given orientation of the magnetic field. Hence the direction of the transport current, flowing along the edge of a sample, coincides with the direction of the paramagnetic current as in case of equilibrium state.

Edge states are populated up to the non-equilibrium electrochemical potential level $\mu_{up}$ so that the empty edge states on the first Landau level are lower than populated bulk states by the value of $\Delta\mu$. The total magnetization current is dominated by the giant diamagnetic current $e^2/h\cdot\Delta\mu$. To explain the experimentally observed giant non-equilibrium magnetization of a 2DEG \cite{Usher2009} it is necessary that $\Delta\mu\geq(20\div60)\hbar\omega_c$, where $\omega_c$ is a cyclotron frequency.

Here we do not discuss the finite magnetic field range of the hysteresis loop as it requires detailed consideration of disorder in nonlinear screening conditions.\cite{Chklovskii1992}

\subsection{Anti-coercive behavior\label{Anti-coercive behavior}}
There are two possible reasons for anti-coercive behavior: (i) an abrupt movement of the boundary of the incompressible liquid at small change of the magnetic field, (ii) topological transitions in quantum Hall liquid. Discuss them in more detail.

(i) at the filling factor $\nu=1$ the magnetic field change $\Delta B$ causes the shift of the boundary of the incompressible liquid relative to the lithographic edge. Besides, the larger the sample the smaller the change of the magnetic field $\Delta B$ is required to shift this boundary by an amount of the order of the depletion width. For example for disk-shaped 2DEG of radius $R$:
\begin{equation}\label{QPC}
\Delta B \approx \frac{W_{depl}}{2R} \cdot B.
\end{equation}

(ii) As seen from Figs.~\ref{fig4}(c) and \ref{fig4}(f) the shift of the boundary of the incompressible liquid leads to topological transitions in quantum Hall liquid. Fig.~\ref{fig4}(c) corresponds to the total transmission of the edge channels through the constriction, while Fig.~\ref{fig4}(f) corresponds to the intense backscattering. Thus, a small change of the magnetic field leads to a drastic change of the magnetoresistance.

Topological transition has been observed in samples of type I: the vanishing of the longitudinal magnetoresistance has been observed indicating total suppression of backscattering in the constriction at down-sweep of the magnetic field, while up-sweep of the magnetic field causes sharp jump of the magnetoresistance from zero to 10~k$\Omega$. Abrupt movement of the boundary of the incompressible electron liquid at small change of the magnetic field is inevitably realized in the both types samples. This is manifested itself in the sharp changes of the magnetoresistance in minor loops measurements.

\begin{figure}
\includegraphics[width=.39\textwidth]{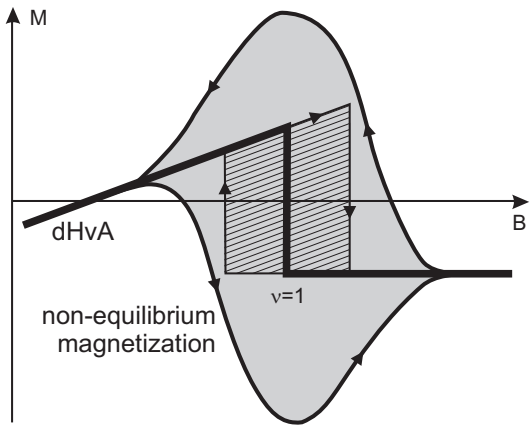}
\caption{Hysteresis of 2DEG magnetization (large loop) in QHE regime against the de Haas - van Alphen oscillations (thick line) is inverted (anti-coercive behavior) that is the response advances the magnetic field sweep. Small loop demonstrates the retarding (coercive) behavior.\label{fig6}}
\end{figure}
The discovered unusual anti-coercive behavior is in accordance with the law of energy conservation. Consider the behavior of the non-equilibrium magnetization of a 2DEG in the QHE regime. In Fig.~\ref{fig6} a typical hysteresis of the magnetization on the background of dHvA oscillations at the filling factor $\nu=1$ is shown. \cite{Matthews2004} The system response (magnetization) also advances the external parameter (magnetic field) change as in the case discussed. However the total work performed by the system  $-\oint{Md\,B}$ (shadow area in Fig.~\ref{fig6}) is negative, that is in accordance with energy conservation law. Retarding response behavior would correspond to the positive total system work (dashed area in Fig.~\ref{fig6}) per cycle contrary to the law of conservation of energy.

\subsection{Comparison with ferromagnetics\label{Comparison with ferromagnetics}}
Let us compare the observed hysteresis of magnetoresistance with the hysteresis of magnetization of ferromagnetics.

It is well known that ferromagnetism is of spin origin. The exchange interaction in ferromagnetics causes the spins align. At the same time, the requirement of a minimum of the total energy of the system leads to its splitting into domains with opposite spin orientations. A change of the magnetic field results in spin flip process near the domain walls causing their ``motion''. However, due to a disorder the transformations of domain structure is retarded with respect to the external magnetic field change. This is the reason of hysteresis with positive coercivity in ferromagnetics. While in a 2DEG a small changes of the external magnetic field lead to sharp motion of the boundary of the incompressible liquid and causes the hysteresis with negative coercivity.

Earlier it has been found \cite{Budantsev2007} that the hysteresis of magnetoresistance is suppressed by temperature. It has been shown that the amplitude of hysteresis linearly increases as the temperature decreases and saturates at certain low temperature ($\sim$400~mK). The non-equilibrium magnetization of a 2DEG has similar temperature dependence.\cite{Matthews2004} The hysteresis of magnetization of ferromagnetic materials is also suppressed by temperature but it has different functional dependence described by Weiss theory of ferromagnetism. \cite{Bertotti1998}

It should be noted that NECs in the QHE regime are induced by sweep of the magnetic field and can be considered as eddy currents as they are often called. However, in ferromagnetics it is enough to reduce the sweep rate to eliminate the influence of eddy currents. In this sense eddy currents in ferromagnetics are time-dependent because of finite longitudinal conductivity $\sigma_{xx}$. While in a 2DEG in the QHE regime when $\sigma_{xx}\approx0$ the generation of NECs is inescapable at any reasonable sweep rate. According to our results NECs relaxation lasts for hours after stopping the magnetic field sweep. In Ref.~[\onlinecite{Kershaw2007}] NECs have been shown to circulate for many hours. In terms of Ref.~[\onlinecite{Bertotti1998}] under conventional experimental conditions NECs can be considered as time-independent.

\section{Conclusions\label{Conclusions}}
We have shown that sweep of the magnetic field in the QHE regime results in drastic changes of electrostatic potential in the bulk with respect to the edge if the inter Landau levels scattering is suppressed. This leads to giant magnetization currents flowing along the edge and topological transitions in a 2DEG with a constriction. The latter results in changes of magnetoresistance.

The hysteresis of magnetoresistance of a 2DEG in the QHE regime has significant phenomenological similarities with the hysteresis of magnetization of ferromagnetic materials, showing multistability, jumps of relaxation, temperature suppression and having the anhysteretic curve. At the same time a fundamental difference expressed in anti-coercive hysteresis behavior of the magnetoresistance has been found.

The possible causes of the limitations of the inter Landau level scattering as well as the possible limiting mechanisms of NECs are beyond the scope of this article and require further investigation. However we have experimentally shown that NECs are induced in a narrow ($\sim$0.5$\mu$m) area along the edge.

Despite the large number of studies devoted to the non-equilibrium state of a 2DEG induced by the sweep of the magnetic field in the QHE regime there is no clear understanding of the phenomena at a microscopic level as well as a systematic phenomenological study of the dependence of the phenomena on parameters of heterostructures and conditions of epitaxial growth up to date. Unfortunately the NECs are insensible in a conventional transport measurement in a linear response regime due to zero longitudinal resistance. In such measurements the condition of the 2DEG is not clear in advance. Such uncertainty can bring unexpected difficulties in studies of a 2DEG in the QHE regime, such as the study of QHE breakdown by the electric current, that has no full and consistent model up to date (see for a review Ref.~[\onlinecite{Nachtwei1999}]). The present study raises the issue of the need to search critical parameters that define the phenomena and explain the phenomena at a microscopic level. Therefore, some existing concepts on the phenomena in 2DEG in the QHE regime can be extended or revised taking into account the found non-equilibrium.

\section*{Acknowledgments\label{Acknowledgments}}
The reported study was supported by RFBR (research project No.14-02-31740-mol-a and No.12-02-00532-a), Program of fundamental scientific research DNIT RAS (project 3.2).


\begin{thebibliography}{99}

\bibitem{Pudalov1984}
V.~M.~Pudalov, S.~G.~Semenchinsky, and V.~S.~Edelman, Sol. St. Commun.\textbf{51}, 713 (1984).
\bibitem{Dolgopolov1992}
V.~T.~Dolgopolov, A.~A.~Shashkin, N.~B.~Zhitenev, S.~I.~Dorozhkin, and K.~von~Klitzing, Phys. Rev. B \textbf{46}, 12560 (1992).
\bibitem{Dolgopolov1993}
V.~T.~Dolgopolov, A.~A.~Shashkin, G.~V.~Kravchenko, S.~I.~Dorozhkin, and K.~von~Klitzing, Phys. Rev. B \textbf{48}, 8480 (1993).
\bibitem{Usher2009}
A.~Usher and M.~Elliott, J. Phys.: Condens. Matter \textbf{21}, 103202  (2009).
\bibitem{Huels2004}
J.~Huels, J.~Weis, J.~Smet, K.~v.~Klitzing, and Z.~R.~Wasilewski, Phys. Rev. B \textbf{69}, 085319 (2004).
\bibitem{Klaffs2004}
T.~Klaffs, V.~A.~Krupenin, J.~Weis, and F.~J.~Ahlers, Physica E \textbf{22}, 737 (2004).
\bibitem{Budantsev2007}
M.~V.~Budantsev, A.~G.~Pogosov, A.~E.~Plotnikov, A.~K.~Bakarov, A.~I.~Toropov, and J.~C.~Portal, JETP Lett. \textbf{86}, 264 (2007).
\bibitem{Budantsev2009}
M.~V.~Budantsev, A.~G.~Pogosov, A.~E.~Plotnikov, A.~K.~Bakarov, A.~I.~Toropov, and J.~C.~Portal, JETP Lett. \textbf{89}, 46 (2009).
\bibitem{Budantsev22009}
M.~V.~Budantsev, A.~G.~Pogosov, A.~K.~Bakarov, A.~I.~Toropov, and J.~C.~Portal, JETP Lett. \textbf{89}, 92 (2009).
\bibitem{Pioro-Ladriere2006}
M.~Pioro-Ladriere, A.~Usher, A.~S.~Sachrajda, J.~Lapointe, J.~Gupta, Z.~Wasilewski, S.~Studenikin, and M.~Elliott, Phys. Rev. B \textbf{73}, 075309 (2006).
\bibitem{Smith2011}
M.~J.~Smith, C.~D.~H.~Williams, A.~Shytov, A.~Usher, A.~S.~Sachrajda, A.~Kam, and Z.~R.~Wasilewski, New J. of Phys. \textbf{13}, 123020 (2011).
\bibitem{Bertotti1998}
G.~Bertotti, \emph{Hysteresis in Magnetism}, (Academic, New York, 1998).
\bibitem{Shikin2002}
V.~B.~Shikin, JETP Lett. \textbf{75}, 465 (2002).
\bibitem{Ruhe2009}
N.~Ruhe, G.~Stracke, Ch.~Heyn, D.~Heitmann, H.~Hardtdegen, Th.~Sch\"{a}pers, B.~Rupprecht, M.~A.~Wilde, and D.~Grundler, Phys. Rev. B \textbf{80}, 115336 (2009).
\bibitem{Matthews2004}
A.~J.~Matthews, K.~V.~Kavokin, A.~Usher, M.~E.~Portnoi, M.~Zhu, J.~D.~Gething, M.~Elliott, W.~G.~Herrenden-Harker, K.~Phillips, D.~A.~Ritchie, M.~Y.~Simmons, C.~B.~Sorensen, O.~P.~Hansen, O.~A.~Mironov, M.~Myronov, D.~R.~Leadley, and M.~Henini, Phys. Rev. B \textbf{70}, 075317 (2004).
\bibitem{Dyakonov1991}
M.~I.~Dyakonov, Solid State Commun. \textbf{78}, 817 (1991).
\bibitem{Eaves1986}
L.~Eaves and F.~W.~Sheard, Semicond. Sci. Technol. \textbf{1}, 346 (1986).
\bibitem{Piazza1999}
V.~Piazza, V.~Pellegrini, F.~Beltram, W.~Wegscheider, T.~Jungwirth, and A.~H.~MacDonald, Nature \textbf{402}, 638 (1999).
\bibitem{DePoortere2000}
E.~P.~De Poortere, E.~Tutuc, S.~J.~Papadakis, and M.~Shayegan, Science \textbf{290}, 1546 (2000).
\bibitem{Jungwirth2001}
T.~Jungwirth and A.~H.~MacDonald, Phys. Rev. Lett. \textbf{87}, 216801 (2001).
\bibitem{Muller1992}
G.~Muller, D.~Weiss, A.~V.~Khaetskii, K.~von~Klitzing, S.~Koch, H.~Nickel, W.~Schlapp, and R.~Losch, Phys. Rev. B \textbf{45}, 3932 (1992).
\bibitem{Khaetskii1992}
A.~V.~Khaetskii, Phys. Rev. B \textbf{45}, 13777 (1992).
\bibitem{Dixon1997}
D.~C.~Dixon, K.~R.~Wald, P.~L.~McEuen, and M.~R.~Melloch, Phys. Rev. B \textbf{56}, 4743 (1997).
\bibitem{Devyatov2004}
E.~V.~Deviatov, A.~W\"{u}rtz, A.~Lorke, M.~Yu.~Melnikov, V.~T.~Dolgopolov, D.~Reuter, and A.~D.~Wieck, Phys. Rev. B \textbf{69}, 115330 (2004).
\bibitem{Kershaw2007}
T.~J.~Kershaw, A.~Usher, A.~S.~Sachrajda, J.~Gupta, Z.~R.~Wasilewski, M.~Elliott, D.~A.~Ritchie, and M.~Y.~Simmons, New J. of Phys. \textbf{9}, 71 (2007).
\bibitem{Ihnatsenka2007}
S.~Ihnatsenka, I.~V.~Zozoulenko, Phys. Rev. B \textbf{75}, 035318 (2007).
\bibitem{Laughlin1981}
R.~B.~Laughlin, Phys. Rev. B \textbf{23}, 5632 (1981).
\bibitem{Chklovskii1992}
D.~B.~Chklovskii, B.~I.~Shklovskii, and L.~I.~Glazman, Phys. Rev. B \textbf{46}, 4026 (1992).
\bibitem{Nachtwei1999}
G. Nachtwei, Physica E \textbf{4}, 79 (1999).


\end{thebibliography}
\end{document}